\documentclass[onecolumn,aps,12pt]{revtex4}
\usepackage{amsmath,graphicx,axodraw}
\def\,{\ifmmode\mskip\thinmuskip\else\leavevmode\thinspace\fi}

\newcommand{\dd}{\mbox{d}}
\newcommand\ba{\begin{eqnarray}}
\newcommand\ea{\end{eqnarray}}
\def\Li#1#2{{\mathrm{Li}}_{#1}\left(#2\right)}
\def\order#1{{\mathcal O}\left(#1\right)}
\def\la{\mathrel{\mathpalette\fun <}}

\def\fun#1#2{\lower3.6pt\vbox{\baselineskip0pt\lineskip.9pt
\ialign{$\mathsurround=0pt#1\hfil##\hfil$\crcr#2\crcr\sim\crcr}}}
\begin{document}

\title{Radiative muon (pion) pair production
in high energy electron-positron annihilation process}

\author{A.B.~Arbuzov}
\affiliation{Joint Institute for Nuclear Research, 141980 Dubna,
Russia}
\author{E.~Barto\v{s}}
\altaffiliation{Department of Theoretical Physics, Comenius
University, 84248 Bratislava, Slovakia.}
\author{V.V.~Bytev}
\affiliation{Joint Institute for Nuclear Research, 141980 Dubna,
Russia}
\author{E.A.~Kuraev}
\affiliation{Joint Institute for Nuclear Research, 141980 Dubna,
Russia}

\date{\today}% It is always \today, today,

\begin{abstract}
Process of muon (pion) pair production with small invariant mass
in the electron--positron high--energy annihilation, accompanied by
emission of hard photon at large angles, is considered. We find
that the Dell--Yan picture for differential cross section is valid
in the charge--even experimental set--up. Radiative corrections both
for electron block and for final state block are taken into account.
\end{abstract}

%\pacs{Valid PACS appear here}
\maketitle

\section{Introduction}
\label{sect1}

Initial state radiation processes in high energy electron-positron
annihilation processes with creation of a hadronic system
provide the laboratory for studying the hadron states created by
virtual photon. Special interest is paid to the case when the
invariant mass of hadron system $\sqrt{s_1}$ is small compared
with the center--of--mass total energy $\sqrt{s}=2\varepsilon$.
Here such interesting physical quantities as the form factor of pion
and nucleon can be investigated. The process of radiative
annihilation into muon and pion pairs, considered here, plays a
crucial role as a normalization one as well as the ones, providing
an essential background process in studying the hadron creation.
Description of its differential cross section with a rather high
level of accuracy (better than $0.5\%$) is the goal of our paper.

We specify the kinematics of radiative muon pair creation process
\begin{eqnarray} \label{process}
&& e_-(p_-)+e_+(p_+)\to\mu_-(q_-)+\mu_+(q_+)+\gamma(k_1),
\end{eqnarray}
as follows:
\begin{eqnarray}
&& \qquad p_\pm^2=m^2,\quad q_\pm^2=M^2, \quad k_1^2=0,
\nonumber \\
&& \chi_\pm=2k_1p_\pm,\quad\chi_\pm'=2k_1q_\pm,\quad s=(p_-+p_+)^2,\quad
s_1=(q_-+q_+)^2,
\nonumber \\
&& t=(p_--q_-)^2,\quad t_1=(p_+-q_+)^2,\quad u=(p_--q_+)^2,\quad
u_1=(p_+-q_-)^2,
\end{eqnarray}
where $m$ and $M$ are the electron and muon (pion) masses, respectively.
Throughout the paper we will suppose
\begin{eqnarray}
s\sim-t\sim-t_1\sim-u\sim-u_1\sim\chi_\pm\sim\chi_\pm'
\gg s_1\sim M^2\gg m^2.
\end{eqnarray}
Situation where $s\gg s_1 \gg M^2$ is allowed too.

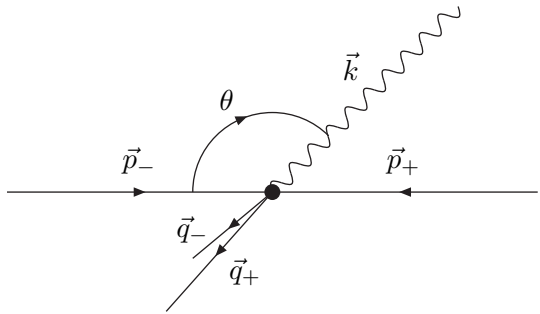
\begin{figure}[htp]
\begin{picture}(300,160)(0,0)
\Vertex(100,70){3} \ArrowLine(0,70)(100,70)
\ArrowLine(200,70)(100,70) \Photon(100,70)(170,140){3}{10}
\Text(150,82)[]{$\vec{p}_+$} \Text(50,82)[]{$\vec{p}_-$}
\Text(130,118)[]{$\vec{k}$} \ArrowLine(100,70)(70,45)
\ArrowLine(100,70)(60,25) \Text(70,56)[]{$\vec{q}_-$}
\Text(90,40)[]{$\vec{q}_+$} \ArrowArcn(100,70)(30,180,45)
\Text(83,105)[]{$\theta$}
\end{picture}
\caption{Kinematical diagram for a radiative event with a small
invariant mass $s_1$.}
\label{kinem}
\end{figure}

We will systematically omit the terms of order $M^2/s$ and $m^2/s_1$
compared with ones of order of magnitude of unity. In $\order{\alpha}$
radiative corrections we drop also terms suppressed by the factor 
$s_1/s$. A kinematical diagram of the process
under consideration is drawn in Fig.~\ref{kinem}.

In this paper we will consider only the charge--even part of 
the differential cross section, which can be measured in an
experimental set--up blind to charges of the created particles.
A detailed study of the charge--odd part of the radiative 
annihilation cross section in general kinematics will be 
presented elsewhere.

Our paper is organized as follows. The next Section is devoted to
the Born--level cross section. Radiative corrections to the final
and initial states are considered in Sect.~\ref{sect3}. In Conclusion
we give some numerical illustration and discuss the resulting precision.
Some details of calculations and useful formulae are given in Appendix.

\section{The Born--Level Cross Section}
\label{sect2}

Within the Born approximation, the matrix element of the
initial state emission process has the form:
\begin{eqnarray}
M_B=\frac{(4\pi\alpha)^{3/2}}{s_1}\bar{v}(p_+)\biggl[
\gamma_\rho \frac{\hat{p}_- - \hat{k}_1+m}{-2p_-k_1}\hat{e}
+ \hat{e}\frac{-\hat{p}_++\hat{k_1}+m}{-2p_+k_1}\gamma_\rho
\biggr] u(p_-) J^\rho
\end{eqnarray}
with
\begin{eqnarray}
J^\rho=\bar{u}(q_-)\gamma^\rho u(q_+)
\end{eqnarray}
for muon pair production, and
\begin{eqnarray}
J^\rho_\pi=(q_- - q_+)^\rho F^{\mathrm{str}}(s_1)
\end{eqnarray}
for the case of charged pions, $F^{\mathrm{str}}(s_1)$ is the pion
strong--interaction form factor.

The corresponding contributions to the cross section is
\begin{eqnarray} \label{born1}
&& \frac{\dd\sigma_B^{j}}{\dd\Gamma}=\frac{\alpha^3}{8\pi^2ss_1^2}R^j,
\qquad R^{j}=B^{\rho\sigma} i^{(0j)}_{\rho\sigma}, \qquad
i^{(0)j}_{\rho\sigma} = \sum J_{\rho}^j({J_{\sigma}}^j)^\star,\quad 
j=\mu, \pi,
\nonumber \\
&& B_{\rho\sigma}= B_gg_{\rho\sigma} + B_{11}(p_{-}p_{-})_{\rho\sigma} +
B_{22}(p_{+}p_{+})_{\rho\sigma},
\nonumber \\
&& B_g=-\frac{(s_1+\chi_+)^2+(s_1+\chi_-)^2}{\chi_+\chi_-}, \quad
B_{11}=-\frac{4s_1}{\chi_+\chi_-},\quad
B_{22}=-\frac{4s_1}{\chi_+\chi_-},
\end{eqnarray}
where we used the short notation $(qq)_{\rho\sigma} = q_{\rho}q_{\sigma}$,
$(pq)_{\rho\sigma} = p_{\rho}q_{\sigma} + q_{\rho}p_{\sigma}$. For the case
with the muon final state 
\ba 
i_{\rho\sigma}^{(0\mu)} =
4\biggl[(q_+q_-)_{\rho\sigma} - g_{\rho\sigma}\frac{s_1}{2}\biggr]. 
\ea 
For the case of pions,
\ba 
i_{\rho\sigma}^{(0\pi)} = |F^{\mathrm{str}}(s_1)|^2(q_--q_+)_{\rho}
(q_--q_+)_{\sigma}. 
\ea

Here we put also the formulae for Born cross section in the case
when only terms of order $(m^2/s)$ are omitted and hard photon are
emitted on large angle $\theta$(see \cite{berends}):
\begin{eqnarray} \label{Born_e}
&& \frac{\dd \sigma_B}{\dd\Omega_-\dd\Omega_\gamma\dd \omega}
=\frac{\alpha^3}{2\pi^2s}\frac{|\vec{q}_+|\omega}
{2\varepsilon-\omega+\omega\cos\theta}A,
\\ \nonumber
&& A=\frac{t^2+t_1^2+u^2+u_1^2}{ss_1}
[W_{\mathrm{even}}+W_{\mathrm{odd}}],
\\ \nonumber
&& W_{\mathrm{even}}=\frac{s}{\chi_+\chi_-}+\frac{s_1}{\chi_+'\chi_-'}
+ \order{\frac{M^2}{s}}\, ,
\\ \nonumber
&& W_{\mathrm{odd}}=
-\frac{t_1}{\chi_+\chi_+'}-\frac{t}{\chi_-\chi_-'}+\frac{u}{\chi_-\chi_+'}
+\frac{u_1}{\chi_+\chi_-'} + \order{\frac{M^2}{s}}\, ,
\end{eqnarray}
for muon pair production, and for pion pair production
\begin{eqnarray}
\frac{\dd\sigma_B}{\dd\Gamma}=\frac{\alpha^3}{4\pi^2s}\frac{tu+t_1u_1}{ss_1}
[W_{\mathrm{even}}+W_{\mathrm{odd}}].
\end{eqnarray}
Functions $W_{\mathrm{even}}$ and $W_{\mathrm{odd}}$ give rise to 
the charge--even and charge--odd contributions to the cross section.
In quantity $W_{\mathrm{even}}$ the term $s/(\chi_+\chi_-)$ is the most 
important in our kinematic region. In what follows, only the even part
will be taken into account.

In our kinematics some simplification takes place:
\begin{gather}
B_g=-2\frac{1+c^2}{1-c^2},\quad
B_{11}=B_{22}=-\frac{16s_1}{s^2(1-c^2)}.
\end{gather}
with
\begin{gather}
c=\cos(\widehat{\vec{p}_-\vec{q}}_-)=\cos\theta.
\end{gather}

We will suppose that the emission angle of hard photon lies
outside the narrow cones around the beam axis $\pi-\theta_0 <
\theta_1 < \theta_0$, with $\theta_0\ll1$, $\theta_0\varepsilon\gg
M$. For initial state radiative process (in paper \cite{KMF} the
scattering channel was considered). The phase volume of the final
particles is
\begin{eqnarray}
\dd\Gamma=\frac{\dd^3q_-}{\varepsilon_-}\frac{\dd^3q_+}{\varepsilon_+}
\frac{\dd^3k_1}{\omega_1}\delta^4(p_++p_--q_+-q_--k_1).
\end{eqnarray}
For the case of small invariant mass of created pair $s_1\ll s$ it
can be rewritten as (see fig.\ref{kinem}):
\begin{gather}
\dd\Gamma=\pi^2\dd x_-\dd c \dd s_1,
\end{gather}
with
\begin{eqnarray}
&& x_\pm=\frac{\varepsilon_\pm}{\varepsilon}, \quad x_++x_-=1,\quad
\omega=\frac{1}{2}\sqrt{s},\quad
\chi_\pm=\frac{s(1\pm c )}{2},
\nonumber \\ \nonumber
&& t=\frac{-sx_-(1-c)}{2} ,\quad
t_1=\frac{-sx_+(1+c)}{2},\quad u=\frac{-sx_+(1-c)}{2},\quad
u_1=\frac{-sx_-(1+c)}{2}\, .
\end{eqnarray}

In the limiting case of small invariant mass of the created pair
omitting terms of the order $s_1/s$ and $M^2/s$,
the Born level cross section takes a simple form:
\begin{eqnarray} \label{born0}
&& \dd\sigma_B^{(\mu)}(p_-,p_+;k_1,q_-,q_+)=
\frac{\alpha^3(1+c^2)}{ss_1(1-c^2)}
\biggl[2\sigma+1-2x_-x_+\biggr]\dd x_- \dd c\; \dd s_1,
\\ \nonumber
&& \dd\sigma_B^{(\pi)}(p_-,p_+;k_1,q_-,q_+)=
\frac{\alpha^3(1+c^2)}{ss_1(1-c^2)}
\biggl[-\sigma+x_-x_+\biggr]\dd x_- \dd c\; \dd s_1,
\\ \nonumber
&& \frac{1}{2}(1-\beta)<x_-<\frac{1}{2}(1+\beta),\quad
\beta=\sqrt{1-\frac{4M^2}{s_1}}, \quad \sigma=\frac{M^2}{s_1}.
\end{eqnarray}
Here $\beta$ is the velocity of a pair component in the center--of--mass
reference frame of the pair.

\section{Radiative corrections}
\label{sect3}

Radiative corrections (RC) can be separated
into three gauge--invariant parts.
They can be taken into
account by the formal replacement (see(\ref{born1})):
\ba \label{zzz}
\frac{R^j}{s_1^2}\longrightarrow
\frac{K^{\rho\sigma}J_{\rho\sigma}^j}{s_1^2|1-\Pi(s_1)|^2}
\ea
where $\Pi(s_1)$ describes the vacuum polarization of the
virtual photon (see Appendix); $K^{\rho\sigma}$ is the initial--state
emission Compton tensor with RC taken into account; 
$J_{\rho\sigma}^j$ is the final state current tensor with 
$\order{\alpha}$ RC. 

First we consider the explicit formulae for RC due to 
virtual, soft, and hard collinear final state emission. 
As concerning RC to the initial state 
for the charge--blind experimental set--up considered here,  
we will use the explicit expression for the Compton
tensor with heavy photon $K^{\rho\sigma}$
calculated in the paper of one of us
\cite{KMF} for the scattering channel and apply the crossing
transformation. Possible contribution due to emission of
an additional real photon from the initial state will be
taken into account too.
In conclusion we give
the explicit formulae for cross section, consider separate the
kinematics of collinear emission, and estimate the contribution of
higher orders of perturbation theory (PT).

\subsection{Corrections to the final state}

The third  part is related with lowest order RC to muon and pion
current \ba J_{\rho\sigma}= i^{(v)}_{\rho\sigma} + i^{(s)}_{\rho\sigma} +
i^{(h)}_{\rho\sigma}. \ea The virtual photon contribution
$i^{(v)}_{\rho\sigma} $ takes into account the Dirac and Pauli form
factors of muon current
\begin{eqnarray}
J^{(v\mu)}_\rho=\bar{u}(q_-)[\gamma_\rho F_1(s_1)+
\frac{\hat{q}\gamma_\rho-\gamma_\rho\hat{q}}{4M}F_2(s_1)]v(q_+),
\quad q = q_+ + q_-, \quad s_1=q^2.
\end{eqnarray}
We have
\begin{eqnarray}
B^{\rho\sigma} i^{(v\mu)}_{\rho\sigma} = B_g \sum|J^{(v\mu)}_{\rho}|^2 +
B_{11}\biggl[\sum|p_-J^{(v\mu)}|^2 + \sum|p_+J^{(v\mu)}|^2\biggr].
%-B_{\rho\sigma}i^{(0\mu)}_{\rho\sigma}
\end{eqnarray}
Here $\Sigma$ means sum over muon spin states and
\begin{eqnarray} \label{ffa}
&& \sum|J^{(v\mu)}_\rho|^2=\frac{\alpha}{\pi}\biggl[-8(s_1+2M^2)f^{(\mu)}
- 12s_1f_2^{(\mu)}\biggr],
\nonumber \\
&& \sum |J^{(v\mu)}p_\pm|^2=\frac{\alpha}{\pi}s^2(1\pm c)(x_+x_-f_1^{(\mu)}
+ \frac{1}{4}f_2^{(\mu)}),
\end{eqnarray}
see Appendix for details. For the pion final state we have \ba &&
B^{\rho\sigma}i_{\rho\sigma}^{(v\pi)} =2 B^{\rho\sigma} i_{\rho\sigma}^{(0\pi)}
 f^{QED}_{\pi},
\nonumber \\
&& B^{\rho\sigma} i^{(0\pi)}_{\rho\sigma} = |F^{\mathrm{str}}(s_1)|^2
\biggl[(4M^2-s_1)B_g + \frac{1}{8}s^2B_{11}(x_+-x_-)^2(1+c^2)\biggr].
\ea
The explicit expression for
the electromagnetic form factor of pion is given in Appendix.

The soft photon correction to the final state currents reads \ba
&& i_{\rho\sigma}^{(s\pi)}
=\frac{\alpha}{\pi}\Delta_{1'2'}i_{\rho\sigma}^{(0\pi)},
\quad\quad\quad i_{\rho\sigma}^{(s\mu)}=
\frac{\alpha}{\pi}\Delta_{1'2'}i_{\rho\sigma}^{(0\mu)},
\nonumber \\
&& \Delta_{1'2'} = - \frac{1}{4\pi}\int\frac{\dd^3k}{\omega}
\biggl(\frac{q_+}{q_+k} - \frac{q_-}{q_-k}\biggr)^2
\bigg|_{\omega\leq\Delta\varepsilon} =
(\frac{1+\beta^2}{2\beta}\ln\frac{1+\beta}{1-\beta}-1)\ln\frac{\Delta\varepsilon^2M^2}{\varepsilon_+\varepsilon_-\lambda^2}
\\ \nonumber && \qquad
+ \frac{1+\beta^2}{2\beta}\biggl[ - g -
\frac{1}{2}\ln^2\frac{1+\beta}{1-\beta} -
\ln\frac{1+\beta}{1-\beta}\ln\frac{1-\beta^2}{4} - \frac{\pi^2}{6}
- 2\Li{2}{\frac{\beta-1}{\beta+1}} \biggr],
\\ \nonumber &&
g=2\beta\int\limits_0^1\frac{\dd t}{1-\beta^2t^2}
\ln\biggl(1+\frac{1-t^2}{4}\frac{(x_+ - x_-)^2}{x_+x_-}\biggr) =
\ln\biggl(\frac{1+\beta}{1-\beta}\biggr)\ln(1 + z -z/\beta^2)
\\ \nonumber && \qquad
+ \Li{2}{\frac{1-\beta}{1+\beta/r}} +
\Li{2}{\frac{1-\beta}{1-\beta/r}} -
\Li{2}{\frac{1+\beta}{1-\beta/r}}
%\\ \nonumber && \qquad
- \Li{2}{\frac{1+\beta}{1+\beta/r}},
\\ \nonumber &&
\beta =\sqrt{1-\frac{4M^2}{s_1}},\qquad z =
\frac{1}{4}\biggl(\sqrt{\frac{x_+}{x_-}} -
\sqrt{\frac{x_-}{x_+}}\biggr)^2, \qquad r =
\left|x_+-x_-\right|,\quad
\Delta=\frac{\Delta\varepsilon}{\varepsilon}. \ea

The contribution of an additional hard photon (with momentum $k_2$) emission by
the muon block, provided $\tilde{s}_{1}=(q_++q_-+k_2)^2\sim s_1\ll s$
can be found by the expression
\begin{eqnarray}
B^{\rho\sigma}i_{\rho\sigma}^{(h\mu)} = \frac{\alpha}{4\pi^2}
\int\frac{\dd^3k_2}{\omega_2}B^{\rho\sigma}\sum J_\rho^{(\gamma)}(J_{\sigma}^{(\gamma)})^*
\bigg|_{\omega_2 \geq \Delta\varepsilon},
\end{eqnarray}
with
\begin{eqnarray}
&& \sum|J^{(\gamma)}_{\rho}|^2=4Q^2(s_1+2k_2q_-+2k_2q_++2M^2)-
8\frac{(k_2q_-)^2+(k_2q_+)^2}{(k_2q_-)(k_2q_+)},
\nonumber \\
&& Q=\frac{q_{-}}{q_-k_2}-\frac{q_{+}}{q_+k_2},
 \nonumber
\end{eqnarray}
and
\begin{eqnarray}
&& \sum|J^{(\gamma)}p_\pm|^2 = - 8Q^2(q_-p_\pm)(q_+p_\pm) + 8(p_\pm
k_2)\biggl(Qq_+\frac{p_\pm q_-}{q_+
k_2}-Qq_- \frac{p_\pm
q_+}{q_-k_2}\biggr)
\nonumber \\ && \qquad +  8(p_\pm k_2)\biggl(\frac{p_\pm q_-}{q_+k_2}+ \frac{p_\pm
q_+}{q_-k_2}\biggr) + 8(p_\pm Q)(p_\pm q_+-p_\pm
q_-)-8\frac{(k_2p_\pm)^2}{(k_2q_+)(k_2q_-)}.
\end{eqnarray}

For the case of charged pion pair production
the radiative current tensor has the form
\begin{gather}
i_{\rho\sigma}^{(h\pi)} = -\frac{\alpha}{4\pi^2}\int
|F^{\mathrm{str}}_\pi(\tilde{s}_{1})|^2 \frac{\dd^3k_2}{\omega_2}
\biggl[\frac{M^2}{\chi_{2-}^2}(Q_1Q_1)_{\rho\sigma} +
\frac{M^2}{\chi_{2+}^2}(Q_2Q_2)_{\rho\sigma}-
\frac{q_+q_-}{\chi_{2+}\chi_{2-}}(Q_1,Q_2)_{\rho\sigma}
\nonumber\\
+g_{\rho\sigma}- \frac{1}{\chi_{2-}}(Q_1,q_-)_{\rho\sigma}
+\frac{1}{\chi_{2+}}(Q_2,q_+)_{\rho\sigma}\biggr]\bigg|_{\omega_2>\Delta\varepsilon},
\\ \nonumber
Q_1=q_--q_++k_2,\qquad  Q_2=q_--q_+-k_2, \qquad \chi_{2\pm} =
2k_2q_{\pm}.
\end{gather}
One can check that the Bose symmetry and the
gauge invariance condition is valid for the pionic current tensor.
Namely it is invariant regarding the permutation of the pion momenta
operation and turns to zero after conversion with 4-vector $q$.

The total sum of RC to the muon current does not depend on
$\Delta\varepsilon/\varepsilon$.

\subsection{Corrections to the intial state}

Let us now consider RC to the Compton tensor with RC, which describe
virtual corrections to the initial state.
In our kinematical region it will be
convenient to rewrite the tensor explicitly extracting large
logarithms. We will distinguish two kinds of large logarithms:
\begin{gather}
l_s=\ln\frac{s}{m^2},\quad l_1=\ln\frac{s}{s_1}.
\end{gather}
We rewrite the Compton tensor (\ref{zzz}) in the form:
\begin{gather}
\label{comp}
K_{\rho\sigma}=(1+\frac{\alpha}{2\pi}\rho)B_{\rho\sigma}+\frac{\alpha}{2\pi}\bigl[\tau_gg_{\rho\sigma}
+\tau_{11}(p_{-}p_{-})_{\rho\sigma}+\tau_{22}(p_{+}p_{+})_{\rho\sigma}-\frac{1}{2}\tau_{12}(p_{-}p_{+})_{\rho\sigma}
\biggr], \\ \nonumber
\rho=-4\ln\frac{m}{\lambda}(l_s-1)-l_s^2+3l_s-3l_1+\frac{4}{3}\pi^2-\frac{9}{2},
\end{gather}
with $\tau_i=a_il_1+b_i$ and \ba a_{11} &=&
-\frac{2s_1}{\chi_+\chi_-}\biggl[\frac{2b^2}{\chi_+\chi_-}+\frac{4s}{a}+
\frac{4(s^2+b\chi_-)}{a^2}-\frac{b^2(2c-\chi_-)}{c^2\chi_-}-\frac{2s+\chi_+}{\chi_+}\biggr],
\\
b_{11} &=& \frac{2}{\chi_+\chi_-}\biggl[-s_1(1+\frac{s^2}{\chi_+^2})G_-
-s_1\biggl(2+\frac{b^2}{\chi_-^2}\biggr)G_+
-\frac{s_1b^2(2c-\chi_-)}{c^2\chi_-}\ln\frac{s}{\chi_+}
\nonumber \\
&-& \frac{s_1}{\chi_+}(2s+\chi_+)\ln\frac{s}{\chi_-}
-\frac{4(s^2+b\chi_-)}{a}-4s-2s_1-\chi_--\frac{b^2}{c} \biggr],
\ea \ba a_{12} &=&-
\frac{2s_1}{\chi_+\chi_-}\biggl[-\frac{4ss_1}{\chi_+\chi_-}
+\frac{8(\chi_+\chi_--s^2)}{a^2}
-\frac{4s}{a}+4ss_1\biggl(\frac{1}{c\chi_-}+\frac{1}{b\chi_+}\biggr)
\nonumber \\
&+& (2ss_1+4\chi_+\chi_-)\biggl(\frac{1}{c^2}+\frac{1}{b^2}\biggr) \biggr],
\\
b_{12}&=&\frac{2}{\chi_+\chi_-}\biggl[\frac{2s_1}{\chi_+^2}(sc-\chi_-\chi_+)G_-
+\frac{2s_1}{\chi_-^2}(sb-\chi_-\chi_+)G_+
\nonumber \\
&+& s_1\biggl(\frac{2ss_1+4\chi_-\chi_+}{c^2}+\frac{4ss_1}{c\chi_-}\biggr)
\ln\frac{s}{\chi_+}
+s_1\biggl(\frac{2ss_1+4\chi_-\chi_+}{b^2}+\frac{4ss_1}{b\chi_+}\biggr)
\ln\frac{s}{\chi_-} \nonumber \\
&+&
\frac{8(s^2-\chi_+\chi_-)}{a}-2s\biggl(\frac{\chi_+}{c}+\frac{\chi_-}{b}\biggr)+2s_1+10s
\biggr], \ea \ba a_g &=&
-2s\biggl(\frac{s_1}{\chi_+\chi_-}-\frac{2}{a}\biggr)
+\frac{c}{\chi_+}\biggl(\frac{3s}{b}-1\biggr)
+\frac{b}{\chi_-}\biggl(\frac{3s}{c}-1\biggr),
\\
b_g &=& -\frac{1}{\chi_+}(\frac{ss_1}{\chi_+}+\frac{2sb+\chi_+^2}{\chi_-})G_-
-\frac{1}{\chi_-}(\frac{ss_1}{\chi_-}+\frac{2sc+\chi_-^2}{\chi_+})G_+
\nonumber \\
&-& \frac{c}{\chi_+}\biggl(\frac{3s}{b}-1\biggr)\ln\frac{s}{\chi_-}
-\frac{b}{\chi_-}\biggl(\frac{3s}{c}-1\biggr)\ln\frac{s}{\chi_+}
+\frac{2s^2-\chi_+^2-\chi_-^2}{2\chi_+\chi_-},
\\ \nonumber
a&=&-(\chi_++\chi_-),\quad b=s_1+\chi_-,\quad c=s_1+\chi_+,
\\ \nonumber
G_-&=&-\ln^2\frac{\chi_-}{s}+\frac{2\pi^2}{3}-2\Li{2}{1-\frac{s_1}{s}}
-2\Li{2}{1+\frac{s_1}{\chi_-}},
\\ \nonumber
G_+ &=& -\ln^2\frac{\chi_+}{s} +
\frac{2\pi^2}{3}-2\Li{2}{1-\frac{s_1}{s}}-2\Li{2}{1+\frac{s_1}{\chi_+}},
\\ \nonumber
&&\tau_{22}(\chi_-,\chi_+)=\tau_{11}(\chi_+,\chi_-).
 \ea

Infrared singularity (the presence of {\em\ photon mass} $\lambda$ in
$\rho$) is compensated by taking into account soft
photon emission from the initial particles:
\begin{gather}
\dd\sigma^{\mathrm{soft}}=\dd\sigma_0\frac{\alpha}{\pi}\Delta_{12},\\
\nonumber \Delta_{12}=\frac{1}{4\pi}\int\frac{\dd^3k}{\omega}
\biggl(\frac{p_+}{p_+k} - \frac{p_-}{p_-k}\biggr)^2
\bigg|_{\omega\leq\Delta\varepsilon}=2(l_s-1)\ln\frac{m\Delta\varepsilon}{\lambda\varepsilon}
+\frac{1}{2}l_s^2-\frac{\pi^2}{3}
\end{gather}
as a result the quantity $\rho$ in formulae (\ref{comp}) will be
\begin{gather}
\rho\to\rho_\Delta=(4\ln\frac{\Delta\varepsilon}{\varepsilon}+3)(l_s-1)-3l_1
+ \frac{2\pi^2}{3}-\frac{3}{2}\, .
\end{gather}

Cross section of two hard photon emission for the case when
one of them is emitted collinearly to the incoming electron or positron
can be obtained by means of the quasi--real electron method~\cite{Baier}:
\begin{eqnarray}
\frac{\dd\sigma_{\gamma\gamma , \,coll}^{j}}{\dd x_-\dd c\;\dd s_1} &=&
\dd W_{p_-}(k_3)\frac{\dd\tilde\sigma_B^j (p_-(1-x_3),p_+;k_1,q_+,q_-)}{\dd x_-\dd c\;\dd s_1}
\nonumber \\ 
&+& \dd W_{p_+}(k_3)\frac{\dd\tilde\sigma_B^j(p_-,p_+(1-x_3);k_1,q_+,q_-)}{\dd x_-\dd c\;\dd s_1},
\end{eqnarray}
with
\begin{eqnarray}
\dd W_p(k_3)=\frac{\alpha}{\pi}[(1-x_3-\frac{x_3^2}{2})
\ln\frac{(\varepsilon\theta_0)^2}{m^2}-(1-x_3)]\frac{\dd x_3}{x_3},
\qquad x_3=\frac{\omega_3}{\varepsilon},\quad
x_3>\frac{\Delta\varepsilon}{\varepsilon}.
\end{eqnarray}
Here we suppose that the polar angle $\theta_3$ between the
directions of the additional collinear photon and the beam axis  does not exceed
some small value $\theta_0\ll 1$, $\ \varepsilon\theta_0\gg m$.

The {\em boosted} differential cross section $\dd\tilde\sigma_B^j(p_-x,p_+y;k_1,q_+,q_-)$
with reduced momenta of the incoming particles reads 
(compare with Eq.~(\ref{born0}))
\begin{eqnarray} \label{boosted}
&& \frac{\dd\tilde\sigma_B^\mu(p_+x_2,p_-x_1;k_1,q_+,q_-)}
{\dd x_-\dd c\dd s_1}=
\frac{\alpha^3(1+2\sigma-2\nu_+\nu_-)(x_1^2(1-c)^2+x_2^2(1+c)^2)}
{s_1sx_1x_2(1-c^2)(x_1+x_2+c(x_2-x_1))},
\nonumber \\
&& \frac{\dd\tilde\sigma_B^\pi(p_+x_2,p_-x_1;k_1,q_+,q_-)}
{\dd x_-\dd c\dd s_1}=
\frac{\alpha^3(\nu_-(1-\nu_-)-\sigma)(x_1^2(1-c)^2+x_2^2(1+c)^2)}
{s_1sx_1^2x_2^2(1-c^2)(x_1+x_2+c(x_2-x_1))},
\nonumber \\
&& \nu_-=\frac{x_-}{y_2},\qquad
y_2=\frac{2x_1x_2}{x_1+x_2+c(x_2-x_1)}.
\end{eqnarray}

In a certain experimental situation, an estimate of the contribution of
the additional hard photon emission outside the narrow cones around the beam axes.
It can be estimated by
\ba
&& \frac{\dd\sigma_{\gamma\gamma , noncoll}^{j}}{\dd x_-\dd c\;\dd s_1}
= \frac{\alpha}{4\pi^2} \int\frac{\dd^3k_3}{\omega_3}
\biggl[\frac{\varepsilon^2+(\varepsilon-\omega_3)^2}
{\varepsilon\omega_3}\biggr] \biggl\{
\frac{1}{k_3p_-}\frac{\dd\sigma_B^j(p_-(1-x_3),p_+;k_1,q_+,q_-)}{\dd x_-\dd c\;\dd s_1} \nonumber
\\ && \qquad
+ \frac{1}{k_3p_+}\frac{\dd\sigma_B^j(p_-,p_+(1-x_3);k_1,q_+,q_-)}{\dd x_-\dd c\;\dd s_1} \biggr\}
\bigg|_{\theta_3\geq\theta_0,\ \
\Delta\varepsilon<\omega_3<\omega_1}, \qquad x_3 = \frac{\omega_3}{\varepsilon}\, .
\ea
It is a simplified expression for the two--photon initial state
emission cross section. Deviation for the case of large angles
emission of our estimation from the exact quantity is small. 
It does not depend on $s$ and slightly depends on $\theta_0$.
For $\theta_0 \sim 10^{-2}$ we have
\begin{eqnarray}
\frac{\pi}{\alpha}\bigg|\frac{\int(\dd\sigma_{\gamma\gamma,\,noncoll}^j-\dd\sigma_{\gamma\gamma,\,
noncoll \, exact}^j)} {\int\dd\sigma^j_B}\bigg| \la 10^{-1}.
\end{eqnarray}

\subsection{Master formula}

By summing all contributions for charge--even set--up we can put the
cross section of radiative production in the form:
\ba \label{all}
&& \frac{\dd\sigma^{j}(p_+,p_-;k_1,q_+,q_-)}{\dd x_-\dd c\;\dd s_1}
= \int\limits_{}^{1}\int\limits_{}^{1} \frac{\dd x_1 \dd x_2}{|1-\Pi(sx_1x_2)|^2}
\frac{\dd\tilde\sigma_B^{j}(p_+x_2,p_-x_1;k_1,q_+,q_-)}
{\dd x_-\dd c\;\dd s_1}
\\ \nonumber && \qquad \times
D(x_1,l_s)D(x_2,l_s)\biggl(1+\frac{\alpha}{\pi}K^{j}\biggr) +
\frac{\alpha}{2\pi}\int\limits_\Delta^1\dd x
\biggl[\frac{1+(1-x)^2}{x}\ln\frac{\theta_0^2}{4}+x\biggr]
\\ \nonumber && \qquad \times
\biggl[\frac{\dd\tilde\sigma^{j}_B(p_-(1-x),p_+;k_1,q_+,q_-)}
{\dd x_-\dd c\;\dd s_1}
+\frac{\dd\tilde\sigma^j_B(p_-,p_+(1-x);k_1,q_+,q_-)}
{\dd x_-\dd c\;\dd s_1}\biggr]
+\frac{\dd\sigma_{\gamma\gamma,\, noncoll}^{j}}{\dd x_-\dd c\;\dd s_1},
\\ \nonumber &&
D(x,l_s)=\delta(1-x)+\frac{\alpha}{2\pi}P^{(1)}(x)(l_s-1)+....,
\\ \nonumber &&
P^{(1)}(x)=\biggl(\frac{1+x^2}{1-x}\biggr)_+,\qquad j=\mu,\pi.
\ea
The boosted cross sections $\dd\tilde\sigma$ is defined above
in Eq.~(\ref{boosted}). The lower limits of the integrals over $x_{1,2}$
depend on experimental conditions.

The structure function $D$ include all dependence on the large
logarithm $l_s$. The so--called $K$-factor reads
\begin{gather}
K^j=\frac{1}{R^j}B^{\lambda\sigma}\left(i_{\lambda\sigma}^{(v)j}+
i_{\lambda\sigma}^{(s)j}+i_{\lambda\sigma}^{(h)j}\right)
%-\frac{3}{2}l_1
+\frac{\pi^2}{3}-\frac{3}{4}+R^{(j)}.
\end{gather}
 It includes the RC which connected with final state
interaction.

Quantities $R^{(j)}$ include the "non-leading" contributions from
the initial state radiation. Generally, they are rather cumbersome
expressions for the case $s_1\sim s$. For the case $s_1\sim M^2\ll
s$ we obtain
\begin{eqnarray}
R^{(\mu)} &=& -\frac{1-c^2}{16(1+c^2)} \bigg\{-2\frac{1+c^2}{1-c^2}
-32\frac{5+2c+c^2}{1-c^2} \ln^2\bigg(\frac{2}{1+c}\bigg)
+4\frac{5-c}{1+c}\ln\bigg(\frac{2}{1+c}\bigg)
\nonumber\\
&+& \frac{2}{(1-2x_-x_+)+2\sigma}
\left[4x_+x_--2\frac{1+c^2}{1-c^2}+4\sigma\right]\bigg\}+(c \to-c),
\\ \nonumber
R^{(\pi)} &=& \frac{1-c^2}{16(1+c^2)} \bigg\{ 4\frac{5+2c+c^2}{1-c^2}
\ln^2\bigg(\frac{2}{1+c}\bigg)
-4\frac{5-c}{1+c}\ln\bigg(\frac{2}{1+c}\bigg)
\\
&+& \frac{1+c^2}{1-c^2}+2 \bigg\}+(c\to -c).
\end{eqnarray}
Here we see the remarkable phenomena: the cancellation of terms
containing $\ln(\frac{s}{s_1})$. In such a way only one kind of
large logarithm $\ln(s/m^2)$. This fact is  the sequence
of renormalization group.

\section{Conclusion}

In paper \cite{KF} the RC to one--photon annihilation of $e\bar{e}$
to hadrons was calculated beyond the leading logarithmic (LL)
approximation. In paper \cite{KMF} the heavy photon Compton tensor
was calculated for the scattering channel. 

We had considered the charge-even contributions, which correspond
to the charge-blind experimental set--up. The charge--odd contribution
to the cross section under consideration appears starting from the
Born level [see Eq.~(\ref{Born_e})]. In the kinematics discussed 
here, the odd part of the cross section is suppressed by the factor
$s_1/s\ll 1$ with respect to the even one.

In papers \cite{Khoze} the inclusive event selection set-up was
considered for radiative pion pair production process. In such a
set-up the hard photon is emitted preferably in direction close to
the beam axes, unlike the kinematics considered here.

We see that cross section in our quasi $2\to2$ kinematics can be
written down in the form of cross--section of Drell--Yan process.
It is not a trivial fact as well as that two types of {\em\ large logarithms}
are present in the problem. Possible background from peripherical
process $e\bar{e}\to e\bar{e} \mu\bar{\mu}$ is negligible in our
kinematics: it is suppressed by factor
$\frac{\alpha}{\pi}\frac{s_1}{s}$ and can be eliminated if
the registration of the primary hard photon (see Eq.~(\ref{process}))
is required by experimental conditions for event selection.

\begin{acknowledgments}
One of us (E.A.K.) is grateful to Slovak Institute for Physics,
Bratislava for hospitality when the final part of this work was
done. This work was supported by RFBR grant 01-02-17437 and INTAS
grant 00366.
\end{acknowledgments}

\section*{Appendix}
\setcounter{equation}{0}
\renewcommand{\theequation}{A.\arabic{equation}}

The one--loop QED form factors of muon and pion are
\begin{eqnarray}
&& \mathrm{Re}\,F^{(\mu)}_1(s_1) = 1
+ \frac{\alpha}{\pi}f_1^{(\mu)}(s_1),\quad
\mathrm{Re}\,F^{(\mu)}_2(s_1)=\frac{\alpha}{\pi}f_2^{(\mu)}(s_1)
\nonumber \\
&& f_1^{(\mu)}(s_1)= \biggl(\ln\frac{M}{\lambda} -
1\biggr) \biggl(1 - \frac{1+\beta^2}{2\beta}l_\beta\biggr) +
\frac{1+\beta^2}{2\beta}\biggl( - \frac{1}{4}l^2_\beta +
l_\beta\ln\frac{1+\beta}{2\beta} + \frac{\pi^2}{3}
\nonumber \\
&& \qquad + 2\Li{2}{\frac{1-\beta}{1+\beta}}
\biggr) - \frac{1}{4\beta}l_\beta ,
\nonumber \\
&& f_2^{(\mu)}(s_1)=-\frac{1-\beta^2}{8\beta}l_\beta,
\nonumber \\
&& \mathrm{Re}\,F^{QED}_{\pi}(s_1) = 1 +
\frac{\alpha}{\pi}f^{QED}_{\pi}(s_1),
\nonumber \\
&& f^{QED}_{\pi}(s_1)=\biggl(\ln\frac{M}{\lambda} - 1\biggr) \biggl(1 -
\frac{1+\beta^2}{2\beta}l_\beta\biggr) +
\frac{1+\beta^2}{2\beta}\biggl( - \frac{1}{4}l^2_\beta +
l_\beta\ln\frac{1+\beta}{2\beta} + \frac{\pi^2}{3}
\nonumber \\ && \qquad
+ 2\Li{2}{\frac{1-\beta}{1+\beta}} \biggr), \qquad
\beta^2 = 1 - \frac{4M^2}{s_1},\qquad
l_\beta=\ln\frac{1+\beta}{1-\beta}\, .
\end{eqnarray}

We put here the expressions for leptonic and
hadronic~\cite{Eidelman:1995ny} contributions into the vacuum
polarization operator $\Pi(s)$:
\begin{eqnarray}
\Pi(s) &=& \Pi_l(s) + \Pi_{h}(s),
\nonumber \\
\Pi_l(s) &=& \frac{\alpha}{\pi}\Pi_1(s)
+ \left(\frac{\alpha}{\pi}\right)^2\Pi_2(s)
+ \left(\frac{\alpha}{\pi}\right)^3\Pi_3(s) + \dots
\nonumber \\
\Pi_h(s) &=& \frac{s}{4\pi\alpha}\Biggl[
\mathrm{PV}\int\limits_{4m_{\pi}^2}^{\infty}
\frac{\sigma^{e^+e^-\to\mathrm{hadrons}}(s')}{s'-s}\dd s'
- \mathrm{i}\pi\sigma^{e^+e^-\to\mathrm{hadrons}}(s)\Biggr].
\end{eqnarray}
The first order leptonic contribution is well known~[1]:
\begin{equation}
\Pi_1(s) = \frac{1}{3}L - \frac{5}{9} + f(x_{\mu}) + f(x_{\tau})
- \mathrm{i}\pi\left[\frac{1}{3} + \phi(x_{\mu})\Theta(1-x_{\mu})
+ \phi(x_{\tau})\Theta(1-x_{\tau})\right],
\end{equation}
where
\begin{eqnarray*}
f(x)\!&=&\! \left\{\begin{array}{l}
\!\!-\frac{5}{9}-\frac{x}{3}+\frac{1}{6}(2+x)\sqrt{1-x}\;\ln\left|
\frac{1+\sqrt{1-x}}{1-\sqrt{1-x}}\right|\ \ \ \,\mathrm{for}\ \ x\leq 1, \\
\!\!-\frac{5}{9}-\frac{x}{3}+\frac{1}{6}(2+x)\sqrt{x-1}\;\arctan\left(
\frac{1}{\sqrt{x-1}}\right)\ \ \ \mathrm{for}\ \  x > 1,\\
\end{array}\right. \\
\phi(x)\!&=&\! \frac{1}{6}(2+x)\sqrt{1-x},\qquad
x_{\mu,\tau} = \frac{4m_{\mu,\tau}^2}{s}\, .
\end{eqnarray*}
In the second order it is enough to take only the logarithmic term
from the electron contribution
\begin{equation}
\Pi_2(s) = \frac{1}{4}(L-\mathrm{i}\pi) + \zeta(3) - \frac{5}{24}\, .
\end{equation}

\end{document}